\documentclass[12pt,a4paper,final]{iopart}

\usepackage{iopams}  
\usepackage{graphicx}

\usepackage[breaklinks=true,colorlinks=true,linkcolor=blue,urlcolor=blue,citecolor=blue]{hyperref}

\begin{document}

\title[Nonlinear power dependence of the spectral properties of an OPO below threshold]{Nonlinear power dependence of the spectral properties of an optical parametric oscillator below threshold in the quantum regime}

\author{Golnoush Shafiee$^{1,2}$, Dmitry V. Strekalov$^{1}$, Alexander Otterpohl$^{1,2,3}$, Florian Sedlmeir$^{1,2}$, Gerhard Schunk$^{1,2,3}$, Ulrich Vogl$^{1,2}$, Harald G.L. Schwefel$^{4,5}$, Gerd Leuchs$^{1,2}$ and Christoph Marquardt$^{1,2}$}
\address{$^1$Max Planck Institute for the Science of Light, Staudtstra{\ss}e 2, 91058 Erlangen, Germany}
\address{$^2$Institute of Optics, Information and Photonics,
Friedrich-Alexander University Erlangen-N{\"u}rnberg, Staudtstra{\ss}e 7 B2, 91058 Erlangen, Germany}
\address{$^3$Erlangen Graduate School in Advanced Optical Technologies (SAOT), Friedrich-Alexander University Erlangen-N{\"u}rnberg,
Paul-Gordan-Stra{\ss}e 6, 91052 Erlangen, Germany}
\address{$^4$The Dodd-Walls Centre for Photonic and Quantum Technologies,
730 Cumberland Street, 9016 Dunedin, New Zealand}
\address{$^5$Department of Physics, University of Otago, 730 Cumberland Street, 9016 Dunedin, New Zealand}
\ead{golnoush.shafiee@mpl.mpg.de}

\begin{abstract}
Photon pairs and heralded single photons, obtained from cavity-assisted parametric down conversion (PDC), play an important role in quantum communications and technology. This motivated a thorough study of the spectral and temporal properties of parametric light, both above the Optical Parametric Oscillator (OPO) threshold, where the semiclassical approach is justified, and deeply below it, where the linear cavity approximation is applicable. The pursuit of a higher two-photon emission rate leads into an interesting intermediate regime where the OPO still operates considerably below the threshold but the nonlinear cavity phenomena cannot be neglected anymore. Here, we investigate this intermediate regime and show that the spectral and temporal properties of the photon pairs, as well as their emission rate, may significantly differ from the widely accepted linear model. The observed phenomena include frequency pulling and broadening in the temporal correlation for the down-converted optical fields. These factors need to be taken into account when devising practical applications of the high-rate cavity-assisted SPDC sources. 
\end{abstract}

\section{Introduction}
Spontaneous parametric down-conversion (SPDC) is a broadband second-order nonlinear optical process in which a high-frequency pump photon spontaneously splits into two lower-frequency photons named signal and idler. In this process energy is conserved. The SPDC efficiency can be enhanced by placing the nonlinear medium in an optical cavity. The resonance property of the cavity limits the bandwidth of the down-converted photons \cite{PhysRevLett.106.053602,PhysRevLett.102.063603,PhysRevLett.83.2556,Fortsch2013}. Cavity-enhanced spontaneous parametric down-conversion has been used to make highly  efficient photon pair sources \cite{PhysRevA.90.063833,doi:10.1063/1.2803761,Wolfgramm:10}. There are two distinct regimes of cavity-assisted parametric down conversion that are typically considered, the parametric down conversion above the OPO threshold and spontaneous parametric down conversion far below the OPO threshold. The threshold is reached when the parametric gain compensates the cavity loss \cite{PhysRevLett.105.263904,PhysRevLett.104.153901}.

The above-threshold OPO is commonly analyzed in a semi-classical manner, that is, neglecting the vacuum fluctuations of the electromagnetic field. Such an analysis shows that no bright parametric light can be generated below the threshold. When the threshold is reached, the zero solution becomes unstable, and macroscopic parametric oscillations spontaneously arise. An above-threshold OPO is similar to a laser, however an OPO works based on the optical gain from parametric amplification in a nonlinear crystal instead of stimulated emission. For the purposes of this paper, we will call the above-threshold OPO regime the classical-nonlinear regime. In this regime, the bandwidths of the down-converted optical fields are narrower than the bandwidth of the cavity, obeying the Schawlow-Townes relation \cite{PhysRev.112.1940,ref10.1051jphys0198900500100120900}.

Below the OPO threshold, there is not sufficient gain to support the oscillations and the role of the resonator is simply to extend the effective length of the parametric crystal by a series of reflections \cite{Garay-Palmett2012}. The resonator also filters the SPDC light, limiting its spectral width to the cavity transmission function. This is a quantum-linear regime, which requires to consider the effect of quantum fluctuations of the vacuum electromagnetic field to occur and is sufficiently accurately described by linear quantum equations of motion. This regime is widely used for the generation of photon pairs \cite{PhysRevA.77.053801,Schunk2015,Luo2015,Pomarico2009,Fortsch2013}.

The transition from the spontaneous to stimulated emission can be considered as a quantum-nonlinear regime when the OPO is operated below the threshold but close enough to it for the nonlinear effects to become important. This regime is not so well studied as the other two but it is very important. In this regime, one can realize a high-rate source of photon pairs, which is interesting for quantum networking \cite{Pomarico2009}, quantum metrology \cite{Wolfgramm2012}, and quantum communication \cite{Zeilinger}.
This regime has been studied with a focus on the signal-idler correlations and squeezing \cite{PhysRevA.30.1386,PhysRevA.32.2887,Graham1968,Fabre1990} as well as for studying the lasing process and super radiance \cite{Kreinberg2017,Wiersig2009,PhysRevA.89.023824}. However, the power dependence of the spectral properties of the parametric photons were not explicitly demonstrated.
Here, we experimentally and theoretically explore this intermediate quantum-nonlinear regime in a monolithic whispering-gallery mode resonator (WGMR) made of lithium niobate, which acts as an optical cavity introducing high parametric gain. The very low absorption in lithium niobate leads to excellent quality factors for WGM resonators on the order of $10^{7}-10^{8}$ \cite{Leidinger:15}, which strongly limits the bandwidth of the parametric process.  WGMRs work based on the principle of the frustrated total internal reflection \cite{Strekalov2016}. We get resonant enhancement of the pump laser and the down-converted photons. The WGMRs are highly tunable \cite{doi:10.1080/09500340.2016.1148211}, low threshold resonators \cite{PhysRevLett.105.263904}, which are versatile sources of heralded single photons \cite{Fortsch2013}. They have been used for single-particle sensing \cite{Arnold2003,Sedlmeir2014}, narrow-band optical filtering \cite{Vyatchanin1992} and lasing \cite{Liang2010}, photon-atom coupling \cite{Schunk2015}, and to mediate various kinds of nonlinear interactions \cite{PhysRevLett.104.153901,Spillane2002}.
In the current work, we study the SPDC in the region below and approaching the OPO threshold of our resonator. WGMRs are specially suitable for this study due to their very low pump threshold. We investigate variations of the emission rate and temporal correlation of the down-converted photons by changing the pump power and going from far below threshold operation to close to threshold operation. Experimental results show a good consistency with our theoretical model.   
This work is arranged as follows. First, we present a theoretical analysis of spectro-temporal properties of narrow band photon pairs generated in cavity assisted parametric conversion in our WGMR. After that, we go through the experimental details and at the end we show the results of the experiment and compare them with our theoretical model.
\section{Theoretical model}
Let us consider a system of three coupled modes: the pump, the signal, and the idler. The Hamiltonian of this system reads
\begin{eqnarray}\label{HPDC}
H &=& \hbar\omega_p(a_p^\dagger a_p+1/2)+\hbar\omega_s(a_s^\dagger a_s+1/2)+\\
&+&\hbar\omega_i(a_i^\dagger a_i+1/2)+ \hbar \kappa(a_pa_s^\dagger a_i^\dagger+a_p^\dagger a_s a_i),\nonumber
\end{eqnarray}
where $a_p$, $a_s$, and $a_i$ are photon annihilation operators and $\omega_p$, $\omega_s$, and $\omega_i$ are eigenfrequencies of the pump, signal, and idler modes, respectively. The parametric coupling rate
\begin{equation}
\kappa= \chi^{(2)}\sigma\frac{\sqrt{(2\pi)^3\hbar\omega_p\omega_s\omega_i}}{n_pn_sn_i} \label{Omega}
\end{equation}
is proportional to the nonlinear susceptibility $\chi^{(2)}$ of the parametric crystal and the factor $\sigma$ describes the three-mode overlap \cite{Strekalov2016}.

Hamiltonian (\ref{HPDC}) leads to a system of three Heisenberg's equations of motion for the frequency-shifted, slowly varying operators $A_p(t) =a_p(t) e^{i\Omega_p t}$, $ A_s(t) =a_s(t)e^{i\omega_s t}$, and  $A_i(t)=a_i(t)e^{i\omega_i t}$. It is convenient to chose the modes eigenfrequencies $\omega_s$ and $\omega_i$ as the central frequencies for the operators $A_s$ and $A_i$, and the external pump frequency denoted $\Omega_p$ for the pump operator $A_p$.  

Our \textit{open} system is characterized by the external pump power $P$ and loss rates $\gamma_p,\,\gamma_s$, and $\gamma_i$ for the pump, signal, and idler modes. Each loss rate $\gamma_j$ consists of an intrinsic dissipative part $\gamma_j^{d}$ and a part $\gamma_j^{c}$ associated with external coupling: $\gamma_j=\gamma_j^{c}+\gamma_j^{d}$ for $j=p,s,i$.  The Heisenberg's equations for an open system are converted to the quantum Langevin equations:
\begin{eqnarray}\label{SlowPDC}
&\frac{d A_p}{dt}&=-(i(\omega_p-\Omega_p)+\gamma_p) A_p-i\kappa A_sA_ie^{i\Delta t}+F_p,\nonumber\\
&\frac{d A_s}{dt}&=-\gamma_sA_s-i\kappa A_pA_i^\dagger e^{-i\Delta t}+F_s,\\
&\frac{d A_i}{dt}&=-\gamma_iA_i-i\kappa A_pA_s^\dagger e^{-i\Delta t}+F_i,\nonumber
\end{eqnarray}
where $\Delta=\Omega_p-\omega_s-\omega_i$ is the detuning of the pump frequency from the parametric resonance. The generalized force operators in the right-hand parts of the System (\ref{SlowPDC}) represent external pump fields as well as thermal and vacuum fluctuations acting on each mode.

To describe PDC below the threshold, we will assume that the pump is not depleted, i.e. neglect the nonlinear interaction term in the first equation of the System (\ref{SlowPDC}). Furthermore, for a strong pump field such as provided by a coherent laser pump, we can treat the operator $A_p$ as a classical amplitude. Now, the first equation is separated from the others and is easily solved for the stationary pump amplitude $\frac{d A_p}{dt}=0$: 
\begin{equation}
A_p=\frac{\langle F_p\rangle}{i(\omega_p-\Omega_p)+\gamma_p}=\sqrt{\frac{2\gamma_p^cP}{\hbar\Omega_p}}\frac{1}{i(\omega_p-\Omega_p)+\gamma_p}.\label{PDCpump}
\end{equation}

Looking for a solution of the two remaining equations of the System (\ref{SlowPDC}) in the spectral form
\begin{eqnarray}\label{Bspectral}
A_{s,i}(t)&=&\int_{-\infty}^{\infty}\,\tilde{A}_{s,i}(\omega)e^{-i\omega t}\frac{d\omega}{2\pi},\nonumber\\
F_{s,i}(t)&=&\int_{-\infty}^{\infty}\,\tilde{F}_{s,i}(\omega)e^{-i\omega t}\frac{d\omega}{2\pi},
\end{eqnarray}
we arrive at
\begin{eqnarray}
(\gamma_s-i\omega)\tilde{A}_s(\omega)&=&\tilde{F}_s(\omega)-i\kappa A_p\tilde{A}^\dagger_i(\Delta-\omega),\nonumber\\
(\gamma_i-i\omega)\tilde{A}_i(\omega)&=&\tilde{F}_i(\omega)-i\kappa A_p\tilde{A}^\dagger_s(\Delta-\omega)\label{Bspectral1}.
\end{eqnarray}
Here, the signal at a frequency $\omega$ is coupled with the idler at the complementary frequency $\Delta-\omega$, and vice versa. This is a manifestation of the frequency entanglement arising in PDC. Solving System (\ref{Bspectral1}) for the signal spectral operator, we obtain
\begin{equation}
\tilde{A}_s(\omega)=\frac{(\gamma_i+i(\Delta-\omega))\tilde{F}_s(\omega)-i\kappa A_p\tilde{F}^\dagger_i(\Delta-\omega)}{Z_s(\omega)},\label{BspectralSol}
\end{equation}
where
\begin{eqnarray}\label{Z1Z2}
Z_s(\omega)&=&(\gamma_s-i\omega)(\gamma_i+i(\Delta-\omega))-\kappa^2|A_p|^2=\\
&=&-\left(\omega-\frac{\Delta}{2}+i\left(\bar{\gamma}+\Upsilon_s\right)\right)\left(\omega-\frac{\Delta}{2}+i\left(\bar{\gamma}-\Upsilon_s\right)\right),\nonumber
\end{eqnarray}
$\bar{\gamma}=(\gamma_s+\gamma_i)/2$, and 
\begin{equation}\label{Ups}
\Upsilon_s^2=\kappa^2|A_p|^2+\left(\frac{\gamma_i-\gamma_s+ i\Delta}{2}\right)^2.
\end{equation}
Similar expressions can be written for the idler.

The product representation of the resonance denominator (\ref{Z1Z2}) indicates the presence of two resonance branches. The  pump-dependent term $\Upsilon_{s,i}$ may have both real and imaginary parts contributing, to the linewidth and central frequency of each resonance branch, respectively. Note that when the pump amplitude turns to zero, System (\ref{BspectralSol}) for $\tilde{A}_{s,i}(\omega)$ collapses to   
\begin{equation}
\tilde{A}_{s,i}^{(0)}(\omega)=\frac{\tilde{F}_{s,i}(\omega)}{\gamma_{s,i}-i\omega}\label{BspectralTriv},
\end{equation}
which describes a linear resonator response to a generalized external force $\tilde{F}_{s,i}(\omega)$. 

To calculate the emission rate of the signal photons $R_s=\langle0|B^\dagger_{s}(t)B_{s}(t)|0\rangle$, we need to convert the intracavity operators $A_s(t)$ to free-space operators $B_{s}(t)$ using the relation 
\begin{equation}
B_{s}(t)=\sqrt{2\gamma_s^c}A_s(t)-B_{s0}(t),\label{in2out}
\end{equation} 
and similarly for the idler. Here, the operator $B_{s0}$ represents external field coupled to the signal mode (in this case vacuum fluctuation). Together with the vacuum fluctuation field of the resonator mode $A_{s0}$, it defines the general force operator:
\begin{equation}
F_s=\sqrt{2\gamma_{s}^c}B_{s0}+\sqrt{2\gamma_{s}^d}A_{s0}.\label{FviaB}
\end{equation}
Substituting (\ref{BspectralSol}) into (\ref{Bspectral}) and using the spectral correlation relations for vacuum fields $\langle \tilde{A}_{j0}(\omega)\tilde{A}^\dagger_{k0}(\omega')\rangle=2\pi\delta_{j,k}\delta(\omega-\omega')$, $\langle \tilde{A}^\dagger_{j0}(\omega)\tilde{A}_{k0}(\omega')\rangle=0$, etc., for $j=s,i$ and $k=s,i$ we find the signal emission rate in a form of a spectral density integral:
\begin{equation}
R_{s}=\int_{-\infty}^{\infty}\,S_{s}(\omega)d\omega,\label{Rout}
\end{equation}
where 
\begin{equation}\label{Sout}
S_s(\omega)=\frac{2\gamma_{s}^c\gamma_i\kappa^2|A_p|^2}{\pi|Z_s(\omega)|^2},
\end{equation}
and similarly for the idler. 

Integration in equation~(\ref{Rout}) can be performed using the method of residues yielding 
\begin{equation}\label{SPDCrate_general}
R_s=\frac{\kappa^2|A_p|^2\gamma_{s}^c\gamma_i\bar{\gamma}}{(\bar{\gamma}^2-\Upsilon^{\prime 2})(\bar{\gamma}^2+\Upsilon^{\prime\prime 2})},
\end{equation}
where $\Upsilon=\Upsilon^{\prime}+i\Upsilon^{\prime\prime}$. If the pump achieves perfect parametric resonance ($\Delta=0$), equation~(\ref{SPDCrate_general}) is reduced to
\begin{equation}
R_s^{\rm res}=\frac{\gamma_{s}^c\gamma_i}{\bar{\gamma}}\frac{P/P_{th}}{1-P/P_{th}},\label{SPDCrate2}
\end{equation}
where the OPO threshold power $P_{th}$ was introduced as \cite{Sturman2012}
\begin{equation}
P_{th}=\frac{\hbar\Omega_p}{\kappa^2}\frac{\gamma_s\gamma_i\gamma_p^2}{2\gamma_{p}^c}\left(1+\frac{\Delta^2}{4\bar{\gamma}^2}\right)\label{Pth0}
\end{equation}
and substituted into $\kappa^2|A_p|^2$.

Expressions  for  the emission  rate  of the idler photons are obtained by
exchanging  the  indices $s$ and $i$ in the above equations. Note that
in general $R_s\ne R_i$. The ratio of the rates for the case of a resonant pump can be found from equation~(\ref{SPDCrate2})  as
\begin{equation}
\frac{R_s^{\rm res}}{R_i^{\rm res}}=\frac{\gamma_s^c}{\gamma_s}\,\frac{\gamma_i}{\gamma_i^c}.\label{Rratio}
\end{equation}
This expression reflects the balance between the emission rates and total loss rates for the signal and idler modes. An identical expression can be derived from a more general equation~(\ref{SPDCrate_general}) in the low-power limit $A\rightarrow 0$.

The result (\ref{Rratio}) is important in the context of using a resonator-assisted SPDC source for absolute calibration of photon counting detectors \cite{1980QuEle..10.1112K,7d043e245d214de6a45243fd401cba63,Brida2006}. This calibration technique is based on the assumption that the signal and idler photons emission rates are equal, which is true for the free-space SPDC, but needs to be appropriately modified for the resonator-assisted SPDC. 

The  emission rates can also be found by converting the coupled-modes Langevin equations into a Fokker-Plank equation and solving it, see e.g. \cite{PhysRevA.28.1560}. This approach allows one to get around the approximation of non-depleted pump and to study the sub- or above-threshold OPO but does not lead to a compact expression for the spectral density. 

In the far below threshold regime, the spectral density (\ref{Sout}) takes on a form consistent with a marginal probability distribution of the joint spectral amplitude for the cavity-assisted PDC described by a well-known expression  \cite{Luo2015,Jeronimo-Moreno2010,Vernon2017}
\begin{equation}
\mathcal{A}_{si}(\omega,\omega^\prime)\propto\frac{\mathcal{A}_p(\omega+\omega^\prime-\Delta)}{(\gamma_s-i\omega)(\gamma_i-i\omega^\prime)},\label{PDCjointSA}
\end{equation}
where $\mathcal{A}_p(\omega+\omega^\prime-\Delta)$ is the spectral amplitude of the pump, $\omega$ and $\omega^\prime$ are frequency detunings from the signal and idler resonance, respectively. 

\begin{figure}[ht]
\begin{center}
\includegraphics[clip,angle=0,height=6cm]{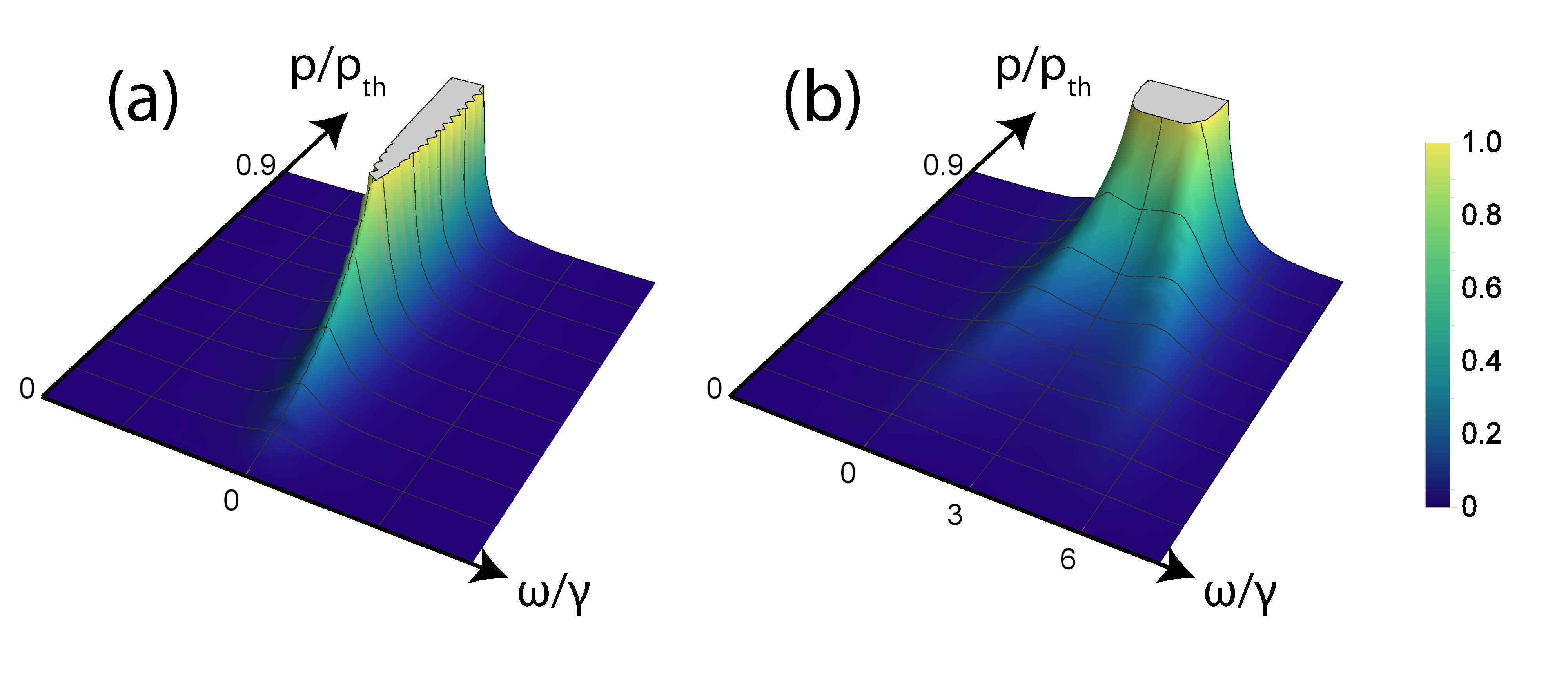}
\caption{Normalized spectral density of the SPDC signal rate to the number of photons for $\gamma_s=\gamma_i\equiv\gamma$ (\ref{Sout}), (a) $\Delta=0$, (b) $\Delta=6\gamma$. }
\label{fig:SPDCspectraOut}
\end{center}
\end{figure}

Our result (\ref{Sout}) for the PDC emission spectrum goes beyond the low-power limit of the previous studies and captures interesting nonlinear phenomena present in parametrically excited resonators. For $\Delta=0$ it shows the spectral line narrowing as the pump power $P$ approaches $P_{th}$ (figure~\ref{fig:SPDCspectraOut}(a)). For $\Delta\ne 0$, the central frequency is being pulled. This is illustrated in figure~\ref{fig:SPDCspectraOut}(b), where we assumed $\gamma_s=\gamma_i\equiv\gamma$ and $\Delta=6\gamma$. In this figure, we see two branches of the signal spectrum, one corresponding to the signal resonance $\omega=0$, the other to the complementary idler resonance $\omega=\Delta$. The idler spectrum has a symmetric shape. The red-shifted signal branch is quantum-correlated (entangled) with the blue-shifted idler branch, and vice versa, so that the frequencies of any signal-idler photon pair always add up to $\Omega_p$.

As the pump power increases, the branches are pulled together, merging when $\kappa^2|A_p|^2$ reaches $(\Delta/2)^2$, i.e.  $\Upsilon =0$. At this point, the frequency offset is  $\omega=\Delta/2$. This corresponds to the oscillation frequency of an above-threshold OPO in presence of parametric detuning \cite{Debuisschert1993}. If the pump power continues to increase, $\Upsilon$ becomes real and begins to contribute to the resonance widths of the down-converted optical fields. For $\Delta=0$, this will occur already at an arbitrarily weak pump and the resonance frequency will always remain at $\omega=0$. Finally, when $\kappa^2|A_p|^2=(\Delta/2)^2+\gamma^2$, which corresponds to the OPO threshold, the bandwidth of the down-converted photons become infinitely narrow and the amplitude infinitely high, which is an artifact of our model neglecting the pump depletion. 

Experimentally, we can measure the temporal correlation between signal and idler photons. To characterize the correlation functions we introduce the Glauber correlation function $G_{i,j}^{(2)}(\tau)=\langle B_{i}^\dagger(t)B_{j}^\dagger(t+\tau)B_{i}(t)B_{j}(t+\tau)\rangle$ and its normalized form $g_{i,j}^{(2)}(\tau)=G_{i,j}^{(2)}(\tau)/G_{i,j}^{(2)}(0)$. Here, the indexes $i$ and $j$ may represent s or i for signal and idler. This leads to the following expressions for the auto- and cross-correlation functions:
\begin{eqnarray}\label{G2ij}
g_{ii}^{(2)}(\tau)&=&1+\frac{\left|\langle B_{i}^\dagger(t)B_{i}(t+\tau)\rangle\right|^2}{\langle B_{i}^\dagger(t)B_{i}(t)\rangle\langle B_{i}^\dagger(t+\tau)B_{i}(t+\tau)\rangle},\nonumber\\
g_{ij}^{(2)}(\tau)&=&1+\frac{\left|\langle B_{i}(t)B_{j}(t+\tau)\rangle\right|^2}{\langle B_{i}^\dagger(t)B_{i}(t)\rangle\langle B_{j}^\dagger(t+\tau)B_{j}(t+\tau)\rangle}.
\end{eqnarray}
Using the spectral representation, we can write
\begin{equation}
g_{ss}^{(2)}(\tau)=1+\left|\frac{\mathcal{G}_{ss}^{(2)}(\tau)}{\mathcal{G}_{ss}^{(2)}(0)}\right|^2\label{g2a}
\end{equation}
with 
\begin{equation}\label{G2a1}
\mathcal{G}_{ss}^{(2)}(\tau)=\int_{-\infty}^{+\infty}\frac{e^{-i\omega\tau}}{|Z_s(\omega)|^2}\,d\omega
\end{equation}
and $\mathcal{G}_{ss}^{(2)}(0)=R_s$. We see that within the framework of our approximation, which neglects pump depletion, signal and idler remain classical, $g_{ss}^{(2)}(\tau)>1$ for any $\tau$, and exhibit thermal auto-correlation statistics, $g_{ss}^{(2)}(0)=2$ for a single mode, for any sub-threshold pump power.

The integral in $\mathcal{G}^{(2)}(\tau)$ can be evaluated using the method of residues, yielding

\begin{eqnarray}
\label{g2b}
\eqalign g_{ss}^{(2)}(\tau)=1+\frac{e^{-2\bar{\gamma}|\tau|}}{4|\Upsilon_s|^2\bar{\gamma}^2}&\left|\right((\bar{\gamma}+\Upsilon_s^\prime)(\bar{\gamma}+i\Upsilon_s^{\prime\prime})e^{\Upsilon_s|\tau|}\nonumber\\
&-(\bar{\gamma}-\Upsilon_s^\prime)(\bar{\gamma}-i\Upsilon_s^{\prime\prime})e^{-\Upsilon_s|\tau|}\left)\right|^2.
\end{eqnarray}
For the cross-correlation, we can write
\begin{equation}
g_{si}^{(2)}(\tau)=1+\frac{\kappa^2|A_p|^2\gamma_{s}^c\gamma_{i}^c}{\pi^2 R_sR_i}\left|\mathcal{G}_{si}(\tau)\right|^2
\label{cross}
\end{equation}
where
\begin{equation}
\mathcal{G}_{si}(\tau)=\int_{-\infty}^{+\infty}\frac{(\gamma_s+i\omega)(\gamma_i+i(\Delta-\omega))+\kappa^2|A_p|^2}{Z_s(\omega)Z_i(\Delta-\omega)}e^{-i\omega\tau}d\omega.
\label{Gcross}
\end{equation}
Again applying the method of residues, we arrive at

\begin{equation}\label{cross_}
\label{cases}
g_{si}^{(2)}(\tau)=1+\frac{4\kappa^2\left|A_p\right|^2\gamma_{s}^c\gamma_{i}^c}{R_s R_i}\cases{\left|r_1+r_2\right|^2,&$\tau>0,$\\
 \left|r_3+r_4\right|^2,&$\tau<0,$\\}
\end{equation}
with
\begin{eqnarray}\label{r1234}
r_1=-i\frac{\left(\gamma_s+\bar{\gamma}+i\frac{\Delta}{2}+\Upsilon_s\right)\left(\frac{\gamma_i-\gamma_s+i\Delta}{2}-\Upsilon_s\right) +\kappa^2|A_p|^2}{2 \Upsilon _s \left(2 \bar{\gamma}-\Upsilon _i+\Upsilon _s\right) \left(2 \bar{\gamma }+\Upsilon _i+\Upsilon _s\right)} e^{-\frac{i \Delta }{2}\tau}e^{-\left(\bar{\gamma }
+\Upsilon_s\right)\tau},\nonumber\\
r_2=i\frac{\left(\gamma_s+\bar{\gamma}+i\frac{\Delta}{2}-\Upsilon_s\right)\left(\frac{\gamma_i-\gamma_s+i\Delta}{2}+\Upsilon_s\right)+\kappa^2|A_p|^2}{2 \Upsilon_s \left(2 \bar{\gamma}-\Upsilon _s+\Upsilon _i\right) \left(2 \bar{\gamma }-\Upsilon _s-\Upsilon _i\right)} e^{-\frac{i \Delta }{2}\tau}e^{-\left(\bar{\gamma }-\Upsilon _s\right)\tau},\nonumber\\
r_3=i\frac{\left(\gamma_i+\bar{\gamma}+i\frac{\Delta}{2}+\Upsilon_i\right)\left(\frac{\gamma_s-\gamma_i+i\Delta}{2}-\Upsilon_i\right)+\kappa^2|A_p|^2}{2 \Upsilon_i \left(2 \bar{\gamma}+\Upsilon _s+\Upsilon _i\right) \left(2 \bar{\gamma }-\Upsilon _s+\Upsilon _i\right)} e^{-\frac{i \Delta }{2}\tau}e^{\left(\bar{\gamma }+\Upsilon _i\right)\tau},\nonumber\\
r_4=-i\frac{\left(\gamma_i+\bar{\gamma}+i\frac{\Delta}{2}-\Upsilon_i\right)\left(\frac{\gamma_s-\gamma_i+i\Delta}{2}+\Upsilon_i\right)+\kappa^2|A_p|^2}{2 \Upsilon_i \left(2 \bar{\gamma}+\Upsilon _s-\Upsilon _i\right) \left(2 \bar{\gamma }-\Upsilon _s-\Upsilon _i\right)} e^{-\frac{i \Delta }{2}\tau}e^{\left(\bar{\gamma }-\Upsilon _i\right)\tau}.
\end{eqnarray}

In the low pump power approximation for $\Delta=0$, equation~(\ref{cross_}) transforms to

\begin{equation}\label{low_cross}
\label{case}
g_{si}^{(2)}(\tau)\approx2+\frac{\gamma_i\gamma_s^c}{\bar{\gamma }R_s}\cases{e^{-2\gamma_s\tau},&$\tau>0$\\
e^{2\gamma_i\tau},&$\tau<0.$\\}
\end{equation}
Low-power limits of the auto- and cross-correlation functions that are discussed in literature, see e.g. \cite{Luo2015}, are consistent with our equations~(\ref{g2b}) and (\ref{cross_}) in the limit when $A_p\rightarrow 0$. Note that even in this case, the auto-correlation function of a parametric light emitted from a resonator is not exponential. 

\section{Experimental setup}
\begin{figure*}[ht]
\centering\includegraphics[width=15cm]{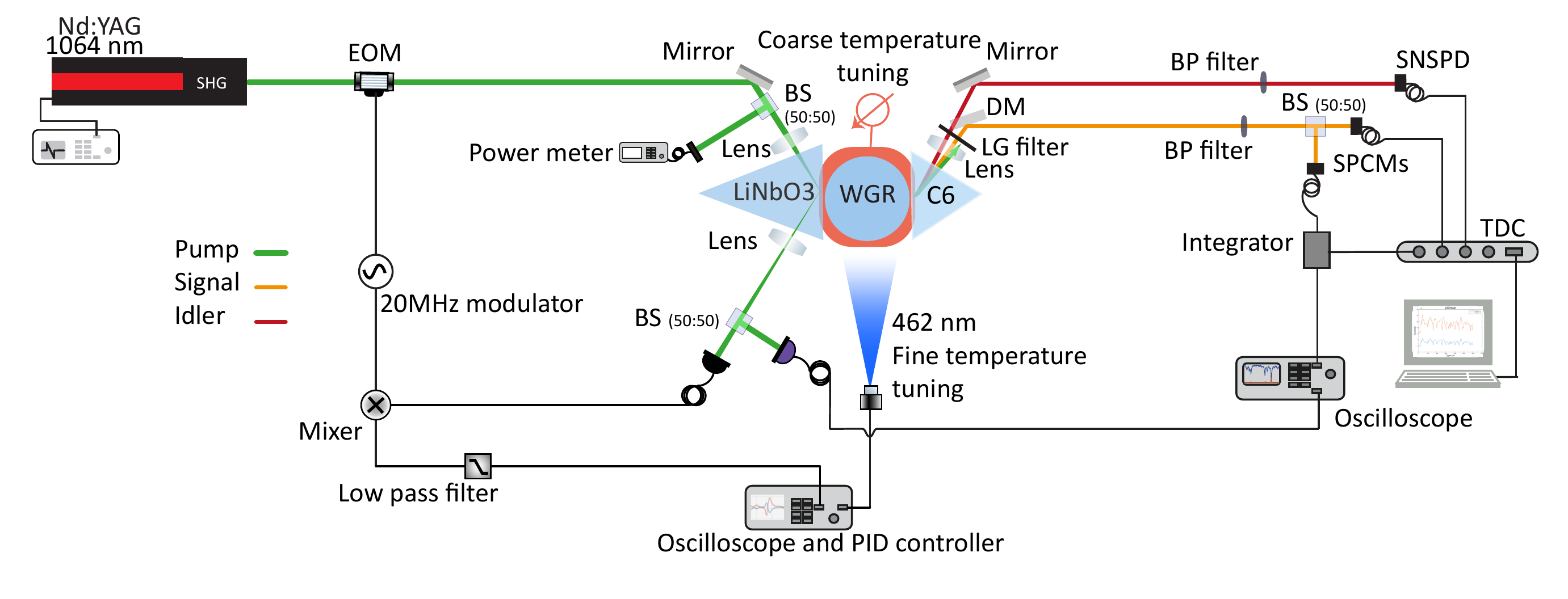}
\caption{Schematic of the experimental setup. Pump light is evanescently coupled to the resonator. The resonator is locked to the laser by the Pound-Drever-Hall locking technique. A Hanbury Brown and Twiss measurement is done on the signal after the 50/50 beam splitter. A home built integrator is used to monitor the signal below the threshold. All three detectors are connected to a time to digital converter (TDC) to register the arrival time of the photons. BS: Beam splitter, DM: Dichroic mirror, EOM: Electro-optic modulator, SHG: Second harmic generation, BP: Band pass filter, C6: Diamond; LiNbO$_3$: Lithium niobate, WGR: Whispering gallery resonator, PID controller: Proportional–integral–derivative controller, SNSPD: Superconducting  nanowire  single  photon  detector, SPCMs: Single photon counting modules, LG filter: Longpass colored glass filter.}
\label{fig:setup}
\end{figure*}

To study the quantum-nonlinear regime, we use a setup schematically shown in figure~\ref{fig:setup}. Our disk-shaped triply resonant WGMR is made of 5\% MgO-doped z-cut LiNbO$_3$.
 For this experiment, we use a WGMR with a disk radius $R  \approx 1.4$ mm and a rim radius $r\approx0.7$ mm, operated at 91$^{\circ}$C. As a pump source, we use  the second harmonic of a continuous wave 1064 nm Nd:YAG laser with kHz-linewidth. We achieve Type-I phase matching for PDC between the extraordinarily polarized pump and the ordinarily polarized down-converted photons. The measured Q-factor of our resonator at critical coupling for 532 nm is $1.4\times10^7$. We use two different prisms to couple the light in and out of the WGMR: an x-cut lithium niobate prism to couple the pump light into the resonator and a diamond prism to couple the down-converted photons out of the resonator. By using two different materials for the couplers, we utilize selective coupling \cite{PhysRevApplied.7.024029,PhysRevApplied.9.024007}, which gives us the possibility to control the  pump incoupling without affecting the coupling rate and bandwidth of the down-converted photons. Both prisms are placed on piezo positioners and controlled with nanometer precision.
 We achieved 50$\%$ coupling contrast for the pump mode of interest at critical coupling. This value is limited by imperfect spatial mode matching between the WGM and the input pump beam. Therefore, it is justified to introduce the incoupled pump power ($\mathrm{p_{in}}$) by multiplying
 the incident pump power with the coupling efficiency of the pump (50$\%$).
 
We use the Pound-Drever-Hall (PDH) technique \cite{Black2000} to lock the resonator pump mode to the pump laser by adjusting the resonator temperature. To achieve that, we generate an error signal from the reflected part of the pump light and feed it to a 462 nm laser diode, which is mounted on top of the resonator and heats it by illumination. Therefore, by detuning the pump laser we can change the temperature of the resonator and achieve resonance operation for down-converted photons \cite{Strekalov2016,Koehler2018}. The speed of the feed back loop is limited by the intrinsic cooling rate of the resonator-coupler system.

After outcoupling the down-converted light, a longpass colored glass filter is used to block the residual pump. Signal and idler are separated by means of a dichroic  mirror. We furthermore use a $960\pm10$ nm band pass filter for the signal and $1200\pm10$ nm band pass filter for the idler. After the filter, the idler is directly guided to a superconducting nanowire single photon detector (Single Quantum Eos SNSPD) and the signal impinges on a 50/50 beam splitter for a Hanbury Brown and Twiss (HBT) measurement \cite{BROWN1956}. Each output arm of the beam splitter is guided to a single photon counting module (Excelitas technologies, SPCM-AQRH-WX-BR).
All three detectors are connected to a time-to-digital converter (qutools, quTAG), which has a digital resolution of 1 ps to record the detection time of incoming photons from all detectors.

We studied the change in the emission rate and the temporal distribution of the down-converted photons for different pump powers going from far below threshold operation to close to threshold operation. The pump laser was locked to its respective mode and the resonator temperature was set to maximize the PDC emission rate. All measurements were done under the same coupling condition which assures the same bandwidth of the cavity for all the measurements. The pump power was measured in front of the incoupling lens as shown in figure~\ref{fig:setup}.

\section{Experimental results and discussions }
We first measured the count rate of the signal photons as a function of the pump power. By tuning the pump laser to reach the maximum count rate for signal and idler, we assumed to have zero parametric detuning. The pump mode was heavily over coupled to increase the threshold in order to render the lock more stable. We measured $\mathrm{FWHM=118}$ MHz for the pump mode during the experiment.

\begin{figure}[ht]
\centering\includegraphics[width=12cm]{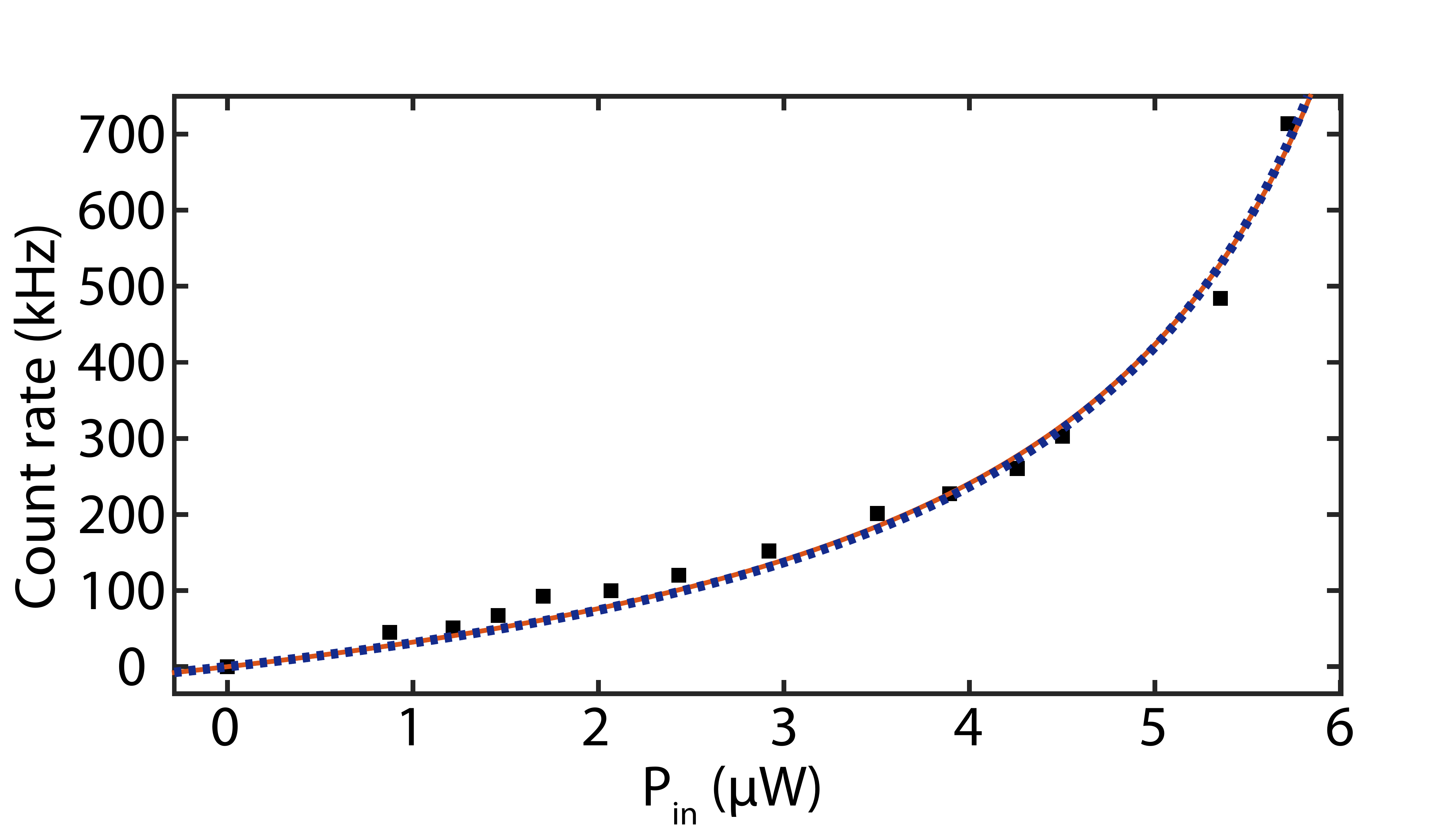}
\caption{Count rate of the signal photons versus the pump intensity. The solid curve is the fit based on the theory (equation~(\ref{SPDCrate2})), the dashed line describes equation~(\ref{SPDCrate2}) including the first order correction (equation~(\ref{frst order corct.eq})).}
\label{fig:cp}
\end{figure}

The result is shown in figure~\ref{fig:cp}.
The solid curve is based on the theoretical model from equation~(\ref{SPDCrate2}). Because of the underlying approximation of a non-depleted pump, equation~(\ref{SPDCrate2}) may not be accurate close to threshold. To evaluate the possible error margin, we substituted solutions for the signal and idler (equation~(\ref{BspectralSol})) into the pump (equation~(\ref{SlowPDC})) and found a first-order correction: 
\begin{equation}
\label{frst order corct.eq}
\frac{P}{P_{th}}\longrightarrow\frac{P}{P_{th}}\left|1-\frac{\kappa^2}{\gamma}\frac{1}{1-P/P_{th}}\frac{1}{i(\omega_p-\Omega_p)+\gamma_p}\right|^2.
\end{equation}
Implementing this correction factor to equation~(\ref{SPDCrate2}) gives the dashed curve in figure~\ref{fig:cp}, which is effectively indistinguishable from the first fit. This gives strong evidence that equation~(\ref{SPDCrate2}) is accurate within the range of our measurements. 
As additional useful information, we can obtain an estimate of the threshold of the system, from this measurement without actually reaching it.
Here, the estimate of the pump threshold based on equation~(\ref{SPDCrate2}) is 7.4 $\mu$W. 


In the theory section, we discussed how the power spectrum of the signal (idler) narrows down by increasing the pump power. Based on our knowledge of the spectral density, we also derived the power dependent auto- and cross-correlation functions (equations~(\ref{g2b}) and (\ref{cross_})) in the time domain. These formulas predict that the correlation time of the parametric photons increases with higher pump power.
To investigate this behavior, we measured the second order correlation function for different pump powers. 

\begin{figure}[ht]
\centering\includegraphics[width=13cm]{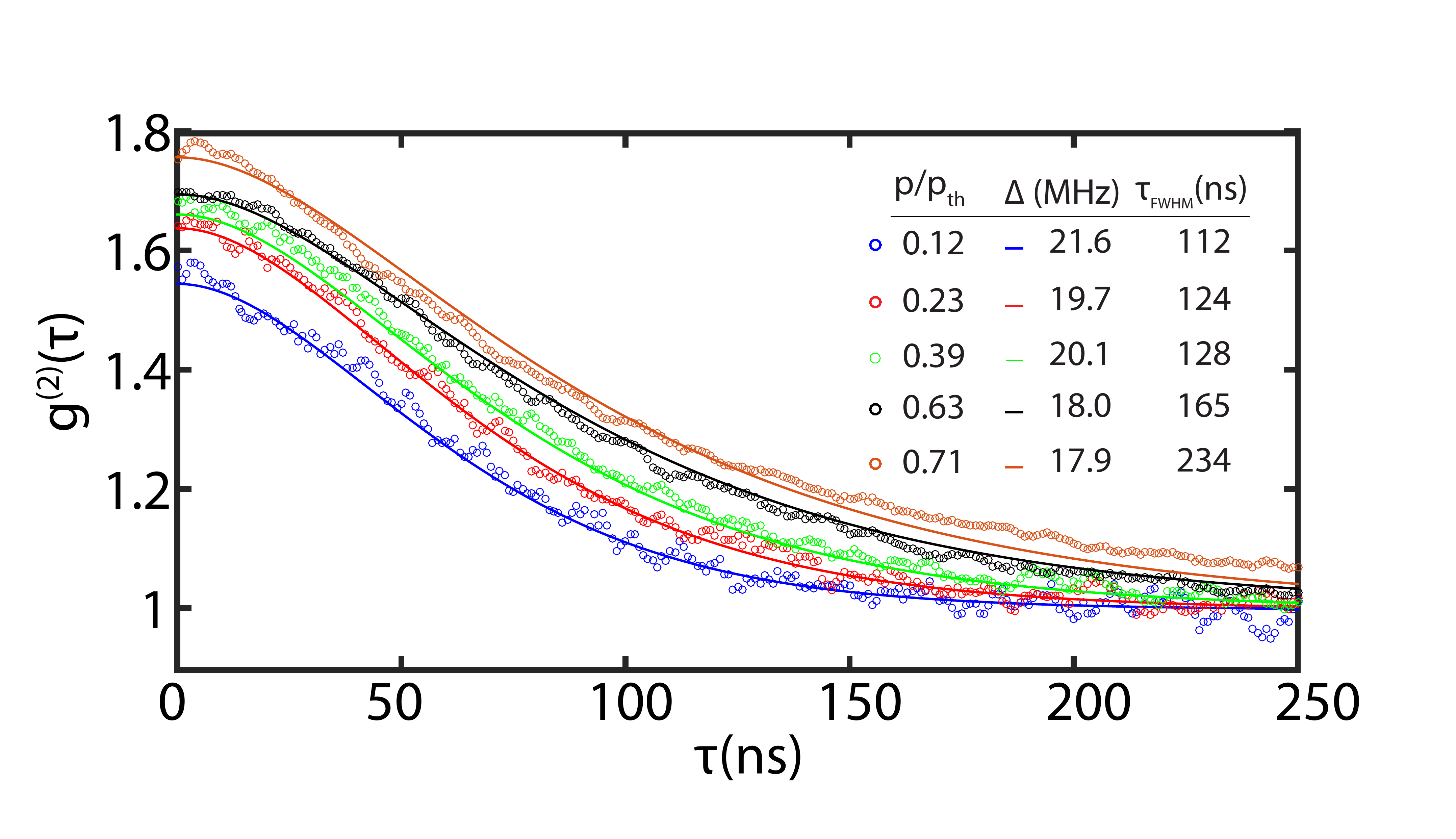}
\caption{HBT measurement of signal photons. The lines are the theoretical fits based on (equation~(\ref{g2b})) for the relevant $p/p_{th}$.} 
\label{fig:g2}
\end{figure}

The results of the HBT measurements are presented in figure~\ref{fig:g2}. Dots represent the experimental data and lines correspond to the theoretical fits obtained from equation~(\ref{g2b}). In order to compare the theory with the experiment, we extracted the bandwidths of the cavity for the down-converted optical fields, ($\gamma_{s,i}$), from the cross-correlation measurement between signal and idler within the low-power regime \cite{Fortsch2013,Luo2015}. We hereby measured the bandwidths of the cavity for signal and idler to be 11 MHz and 16 MHz, respectively.

By fitting the experimental data with equation~(\ref{g2b}), we can find not only the correlation time but also the detuning of the pump frequency from the parametric resonance $(\Delta)$ for each measurement. As described in the previous section, the resonator pump mode is locked to the pump laser but the parametric fields frequencies can freely move within the bandwidth of the pump mode due to thermal drifts \cite{Strekalov2016}. The technical implementation of the locking scheme caused the reported systematic PDC detuning of the order of 20 MHz. This detuning is much smaller than the phase matching width which is dominated by the pump WGM  linewidth of 118 MHz.

 
In the low-power regime, the photon statistics of the signal and idler photons are expected
to show a characteristic bunching peak typical for the thermal distribution: $g_{ii}^{(2)}(0)=1+1/k$, with k being the number of effective modes \cite{Christ2011,Eberly2006,Ivanova2006}. In the case of a single-mode system, one can measure $g_{ii}^{(2)}(0)=2$ below threshold \cite{Fortsch2013}. Here, for $p/p_{th}=0.12$, we measure $g_{ss}^{(2)}(0)\approx1.6$, which means we approximately have 1.7 modes in our system. The non-integer number of modes can be understood as the presence of more than one mode in the system, which are not equally excited \cite{Fortsch2013}. Our results show an increase of $g_{ss}^{(2)}(0)$ by increasing the pump power (from far below the threshold to close to the threshold), which means that one of the excited modes becomes more dominant in the system and suppresses the other weekly exited modes. 
This result is consistent with the so called pump clamping behaviour \cite{2015arXiv150405917N,Otterpohl2019}, although this behavior hinges on the pump depletion and should not arise within our approximation.

If we approach the threshold even closer by further increasing the pump power, the peak value of the auto-correlation function quickly drops, reaching unity at the threshold and one can observe $g_{ss}^2(\tau) \approx 1$ for all $\tau$. This transition from thermal to coherent photon statistics is associated with the pump depletion and cannot be described within our model.

\begin{figure}[ht]
\centering\includegraphics[width=13cm]{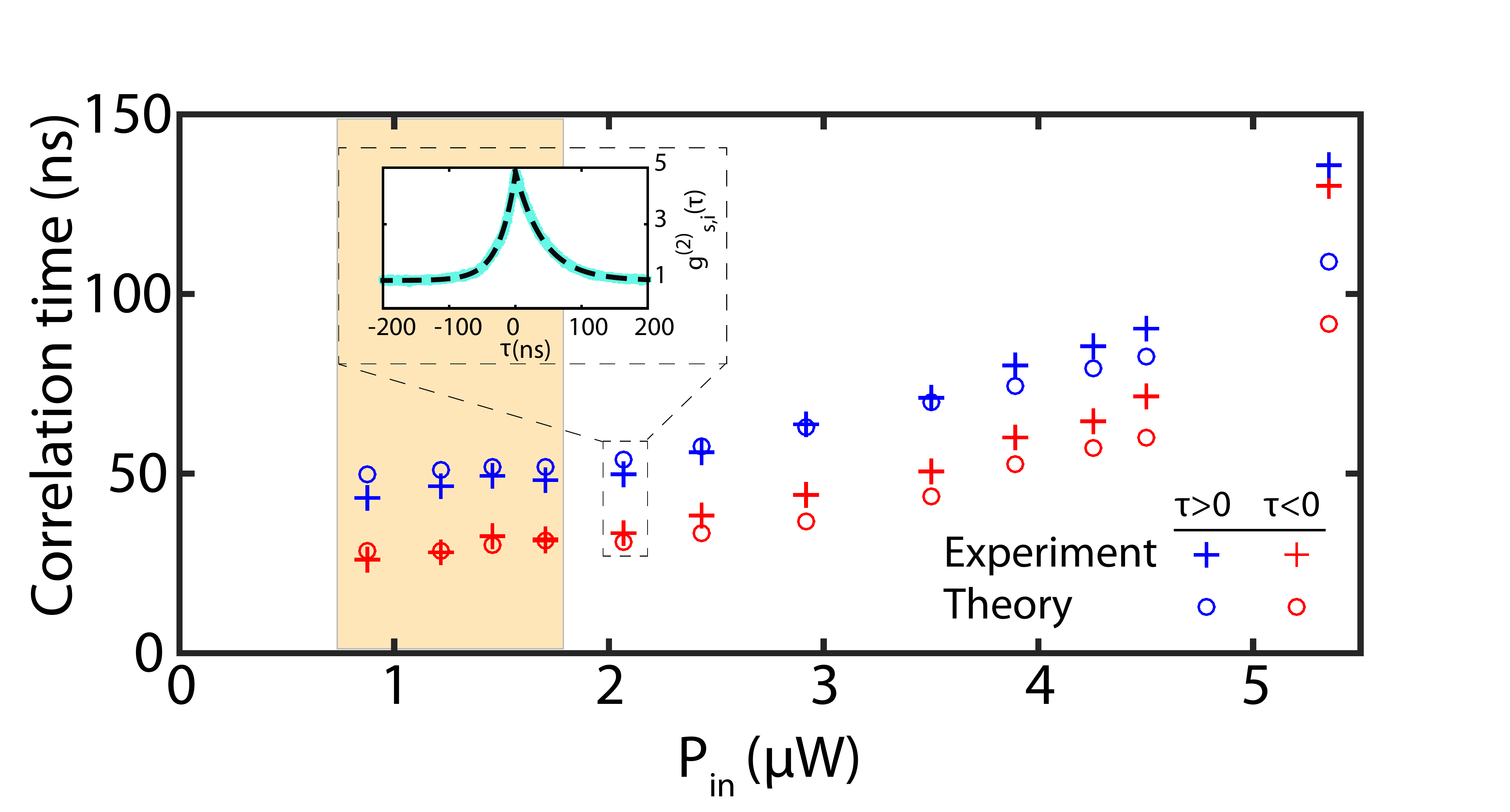}
\caption{Cross-correlation time of the down-converted optical fields versus the pump intensity. The inset shows a normalized cross-correlation function. The dashed line shows the exponential fit $e^{(\pm t/\tau_{s,i})}$, which $\tau_{s,i}$ is the correlation time for signal (idler). Theoretical values are base on equation~(\ref{cross_}) by considering parametric detuning, which was extracted from the theory fits on the auto-correlation data for each measurement. The shaded area shows the low-power regime, in which the correlation time is equal to ${1/(2\gamma_{s,i})}$.}
\label{fig:Bp}
\end{figure}

We also studied the behavior of the cross-correlation between signal and idler for different pump powers. Figure~\ref{fig:Bp} shows the variation of the cross-correlation time of the down-converted optical fields by increasing the pump power. With our knowledge regarding the parametric detuning for each measurement, we used equation~(\ref{cross_}) to find the theoretical prediction for the cross-correlation. The experiment shows a stronger effect of nonlinear linewidth narrowing (correlation time growth) than the theory. This can be because our theory only takes into account the first-order effects, neglecting the pump depletion and its secondary effects on the signal and idler properties. 

In the low-power regime, the signal-idler correlation function, $g^{(2)}_{si}(\tau)$, is known to be asymmetric \cite{Luo2015} as shown in the inset in figure~\ref{fig:Bp}. The two exponential slopes correspond to the correlation times ($\tau_{s,i}$) of the down-converted optical fields. For the low-power measurements (the shaded region in figure~\ref{fig:Bp}), the correlation time is equal to $1/(2\gamma_{s,i})$.
 As the pump power increases, the spontaneous emission is amplified and the correlation time of the down-converted fields increases and and consequently deviates from the inverse of the bandwidth of the cavity. Furthermore, as we approach the OPO threshold, the difference between the correlation time of these optical fields is getting smaller, while the correlation function becomes symmetric. This illustrates that in the low-power regime, the cavity governs the physical properties of the parametric photons while close to the threshold in the quantum nonlinear regime, the nonlinear optical effects prevail. This behaviour is  similar to the lasing process which is shown in \cite{Kreinberg2017}.
 
This result also enables us to find the limit for the pump power defining the low-power regime, where the correlation times of the parametric optical fields are not affected by the pump power. This regime is essential for single photon experiments.

\section{Conclusion }
In this paper, we presented a theory for the cavity-assisted parametric
down conversion process below but close enough to the OPO threshold such that the stimulated processes become important. The theoretical results for the emission rate and the second order correlation of the down-converted optical fields as functions of the pump power are supported by our experimental data. 
These results can be important for various cavity assisted PDC experiments below the oscillation threshold, which need to operate in specific pump power regimes in order to meet the expectations regarding the spectral and statistical properties of the PDC light that are often taken for granted. Conversely, operating cavity-assisted PDC in the nonlinear-quantum regime opens up the possibilities for engineering such properties to meet specific needs of quantum optics applications.


\section{Acknowledgment }
We thank Dr. Valentin Averchenko for useful discussions. This project has partially received funding from the European Commission Horizon 2020 research and innovation programme under the Future and Emerging Technologies Open Grant Agreement Super-Pixels No. 829116.

\section*{References}
\bibliography{ref}

\end{document}